\documentclass{JHEP3}
\usepackage{amsmath,amssymb,amsfonts,mathrsfs}
\usepackage{url}
\usepackage{graphicx}
\usepackage[all]{xy}

\newcommand{\C}{\mathbb{C}}
\newcommand{\R}{\mathbb{R}}
\newcommand{\Z}{\mathcal{Z}}
\newcommand{\calO}{\mathcal{O}}
\newcommand{\bP}{\mathbb{P}}

\title{M theory and the Coulomb phase of higher rank DT invariants}

\author{Haitao Liu\\
  Department of Mathematics and Statistics,\\
  University of New Brunswick, Fredericton, Canada, E3B 5A3\\
  Email: \email{haitao.liu@unb.ca}\\}

\abstract{In this paper, we advance an M theory model corresponding to the Coulomb phase of higher rank Donaldson-Thomas(DT) invariants.} 


\preprint{}

\begin{document}
\section{Introduction}
Counting BPS states, especially counting the bound states of D-branes wrapped around cycles of the internal Calabi-Yau threefold, is an extremely important problem both in string theory and in algebraic geometry. In string theory, it was found that the indices (or degeneracies) of these BPS states are related to the entropy of a family of supersymmetric black holes, degeneracies of instantons in certain gauge theories, topological string theory, statistical models of crystal melting \cite{Strominger:1996sh, Sen:2007qy, Ooguri:2004zv, Okounkov2007, Iqbal:2003ds, Cirafici:2008sn, Sulkowski:2009rw, Szabo:2009vw}. In \cite{Ooguri:2004zv}, the authors conjectured a formula relating to the black hole entropy and partition function of topological strings--the Ooguri-Strominger-Vafa(OSV) conjecture. Mathematically, the Donaldson-Thomas(DT) invariants can be related to the degeneracy of D0 and D2-branes bound to a single D6-brane in type IIA theory \cite{Iqbal:2003ds, Maulik2007}. 

To date, much has been revealed, including the wall-crossing formula of the DT invariants with a single D6-brane wrapped on certain special Calabi-Yau threefolds \cite{Iqbal:2003ds, Maulik2007, Maulik2007a, Cirafici:2008sn, Aganagic2009}. But for the higher rank DT invariants, which are known to be the degeneracies of D0 and D2-branes bound to multiple D6-branes wrapped on Calabi-Yau threefolds, we only know few details \cite{Nagao2010, Chuang:2010wx, Stoppa2009, Toda2009, Cirafici:2008sn}. Most of these papers aim to investigate the bound states of the D6 branes, in contrast to the analysis here, where we generalize Aganagic, et. al.'s results in \cite{Aganagic2009} to the N separated D6-branes case. This is known as the Coulomb phase of the higher rank Donaldson-Thomas invariants. Comparing with the bound states of the D6 branes in \cite{Nagao2010, Chuang:2010wx, Stoppa2009, Toda2009}, the Coulomb phase is the other exteme, with the D6 branes not binding at all to each other.\footnote{In \cite{Stoppa2009}, the author mentioned that  the Coulomb phase is a limit of their results when the central charge becomes degenerate.} Furthermore, since the D6 branes do not bind, then it is not hard to imagine that there should be the product structure which was discovered in \cite{Cirafici:2008sn}. This product structure is needed, especially since  the M-theory derivation presented here makes it very clear that an extreme limit of the geometry is required for the results here to hold.

In Section 2, we will briefly summarize the results in \cite{Aganagic2009}. In Section 3 we will generalize \cite{Aganagic2009} to the multi D6-brane case and propose a formula corresponding to the Coulomb phase of higher rank DT invariants. In Section 4 we will check two simple examples--$\C^3$ and the resolved conifold. In Section 5 we will describe some future work. 

\section{Preliminaries}

In this section we will review the work in \cite{Aganagic2009}. Suppose there is a single D6 brane wrapped on a Calabi-Yau threefold $X$. We want to calculate the BPS partition function of this D6 brane bound to arbitrary number of D2 and D0 branes. First, we lift the D6 brane to the Taub-Nut space via \cite{Gaiotto:2005gf}. The Taub-Nut space is an $S^1$-bundle over $\R^3$, where the $S^1$ collapses at the origin in $\R^3$. Notice that the Taub-Nut space is asymptotic to $\R^3\times S^1$. We denote the radius of this Taub-Nut circle $S^1$ by $R$. If $R$ goes to infinity, then the Taub-Nut space becomes $\R^4$.  Further, the D6 brane is placed at the origin. Thus we get a five dimensional space whose spacial part is the Taub-Nut space. So counting certain BPS states in Taub-Nut space will correspond to the original counting problem. In order to make this correspondence achievable, we need to make the following two assumptions:
\begin{enumerate}
\item We require the background moduli of the CY space as well as the chemical potential to be chosen so that M2-branes wrapped in various way have parallel central charge;
\item The only BPS states in 5D are particles. In other words, we restrict the CY space to the CY space without any compact four-cycles.
\end{enumerate}
The first assumption can be achieved by considering the vanishing K\"{a}hler parameters of the CY and at the same time turning on the B field flux through two-cycles of Calabi-Yau in type IIA. Thus for a state generated by D2 branes wrapping a class $\beta$ in the Calabi-Yau the central charge is 
\begin{equation}
Z(n, \beta)=\frac{1}{R}(n+B\cdot \beta),
\end{equation}
where $n$ is the total momentum along the Taub-Nut circle and $R$ is the radius of the Taub-Nut circle. The second assumption guarantees that there are no string-like states coming from M5 branes wrapped on the compact four-cycles. 

We know that if we fix $B$, then the degeneracies of M-theory don't change when varying the Taub-Nut circle radius $R$, i.e. no states decay during this procedure. So we take $R\rightarrow \infty$ with the result that the Taub-Nut space becomes $\R^4$. Thus the 11 dimensional space in M-theory becomes 
$$ X\times \R^4\times S_t^1,$$ 
where $S^1_t$ is the Euclidean time circle. Now we would like to count the particle degeneracy on $\R^4\times S_t^1$. For each 5D particle, there is  a field $\Phi(z_1, z_2)$ on $\R^4$, where $z_1, z_2$ are two complex coordinate of $\R^4$. The ground-state wave functions of the particle with different momenta correspond to the modes $\alpha_{l_1,l_2}$ in the decomposition of $\Phi(z_1, z_2)$: 
$$
\Phi(z_1, z_2)=\sum_{l_1,l_2}\alpha_{l_1,l_2}z_1^{l_1}z_2^{l_2}.
$$
We know that the $U(1)\in SU(2)_L$ charge of $z_1,z_2$ is 1. Further, under $SU(2)_R$ there are $N_\beta^{m, m'}$ 5D states of intrinsic spin $(m, m')$. Thus the degeneracy we need is the Gopakumar-Vafa invariants:
$$
N_\beta^m=\sum_{m'}(-1)^{m'}N_\beta^{m, m'}.
$$
Hence the total angular momentum is 
$$
n=l_1+l_2+m.
$$
Then the BPS states degeneracy in a chamber specified by $R$ and $B$ is expressed by 
\begin{eqnarray}
\Z(R, B)&=&\text{Tr}_{\text{Fock}}q^{Q_0}Q^{Q_2}|_{Z(n,B)>0}\nonumber\\
&=& \prod_{\beta, m}\prod_{l_1+l_2=n}(1-q^{l_1+l_2+m}Q^\beta)^{N_\beta^m}|_{Z(n,B)>0}\nonumber\\
&=& \prod_{\beta, m}\prod_{n=1}^\infty(1-q^{n+m}Q^\beta)^{nN_\beta^m}|_{Z(n,B)>0}.
\end{eqnarray}
When $R>0, B\rightarrow\infty$ for all K\"{a}hler classes, it follows that  
\begin{equation}
\Z(R>0, B\rightarrow\infty)=\Z_{\text{DT}}=M(q)^{\chi/2}\Z_{\text{top}}(q, Q),
\end{equation}
where $M(q)=\prod_n(1-q^n)^{-n}$ is the MacMahon function and $\chi$ is the Euler characteristic of the Calabi-Yau space $X$. 

\section{Coulomb phase of higher rank DT invariants on multi D6-branes}

Suppose we have N D6-branes wrapped on the Calabi-Yau threefold $X$. Then on $X$ we may construct a U(N) cohomological gauge theory \cite{Iqbal:2003ds, Cirafici:2008sn, Szabo:2009vw}. If $X=\C^3$, then the instantons also have the ADHM structure \cite{Moore:1998et, Cirafici:2008sn}. Thus we can use the equivariant cohomology technique to construct a quiver matrix model and calculate the contribution of instantons \cite{Iqbal:2003ds, Jafferis:2007sg, Cirafici:2008sn, Awata:2009dd}. From \cite{Gaiotto:2005gf} we know that the N D6-branes can lift to M-theory by putting these D6-branes on the origin of the Taub-Nut space. Thus when the radius $R$ of the Taub-Nut circle goes to infinity, we will get $\R^4/\mathbb{Z}_N$ \cite{Gaiotto:2005gf, Dabholkar2008}. Since we only focus on the Coulomb phase, we need to use the multi-centered Taub-Nut space to lift D6-branes instead of the single centered Taub-Nut space. The metric of multi-centered Taub-Nut space is
\begin{equation}
ds^2=V^{-1}(dx^{11}+\vec{\omega}\cdot d\vec{x})^2+Vd\vec{x}\cdot d\vec{x}, \label{multi-taub-nut}
\end{equation}
where $\vec{x}=(x, y, z)$ are coordinates in $\R^3$. The harmonic function $V(\vec{x})$ and the vector potential $\vec{\omega}$ are given by 
\begin{eqnarray}
&&V=1+\sum_{i=1}^N\frac{R}{|\vec{x}-\vec{x_i}|}, \\
&&\vec{\omega}=\sum_{i=1}^N\vec{\omega}_i,\\
&&\vec{\omega}_i\cdot d\vec{x}=\frac{z-z_i}{|\vec{x}-\vec{x}_i|}\frac{(x-x_i)dy-(y-y_i)dx}{(x-x_i)^2+(y-y_i)^2},
\end{eqnarray} 
where $\vec{x}_i=(x_i, y_i, z_i), i=1, \cdots, N$ are the locations of N D6-branes \cite{Ida:2007vi}. When $\vec{x}\rightarrow \vec{x}_i$, under the spherical coordinate $r, \theta, \phi$ defined by $r:=|\vec{x}-\vec{x}_i|, x-x_i=r\cos{\theta}\sin{\phi}, y-y_i=r\cos{\theta}\cos{\phi}, z-z_i=r\sin{\theta}$, the metric (\ref{multi-taub-nut}) reduces to
\begin{equation}
ds^2=\frac{R}{r}(dr^2+r^2d\theta^2+r^2\sin^2{\theta}d\phi^2+r^2(dx^{11}+\cos{\theta}d\phi)^2),
\end{equation}
which is a flat metric on $\R^4$ \cite{Gaiotto:2005gf}. Here $\vec{x}\rightarrow \vec{x}_i$ is not equivalent to $R\rightarrow \infty$ as is the case for the single-centered Taub-Nut space. When $R\rightarrow \infty$, the multi-Taub-Nut metric reduces a flat metric on $\R^4/\mathbb{Z}_N$ \cite{Witten1997b, Gaiotto:2005gf}. Further, as $R\rightarrow \infty$, the centers $\vec{x}_i, i=1, \cdots, N$ will collapse together and the gauge group will be enhanced from $U(1)^N$ to $U(N)$, which implies that we cannot remain in the Coulomb phase. Thus the degeneracy of BPS states of the Coulomb phase would be changed as $R\rightarrow \infty$. Since we only focus on the Coulomb phase, so here we will keep $R$ fixed or perturb $R$ slightly in order to remain in the Coulomb phase. Thus when we approach each center we are able to associate an $\R^4$ with that center. Since we have $N$ centers $\vec{x}_i$, then we can get $N$ $\R^4$s. Hence the space we are interested is 
\begin{equation}
(\R^4)^{\otimes n}:=\underbrace{\R^4\otimes\cdots\otimes\R^4}_N.
\end{equation}

Now we denote by $z_{i,1}, z_{i,2}$ the complex coordinates of the $i^{th}$ $\R^4$. Then for each 5D particle we get a field 
\begin{equation}
\phi_i(z_{i,1}, z_{i,2})=\sum_{l^i_1,l^i_2}\alpha_{l^i_1, l^i_2}(z_{i,1})^{l^i_1}(z_{i,2})^{l^i_2},
\end{equation}
in the $i^{th}$ $\R^4$. As in \cite{Aganagic2009}, the modes $\alpha_{l^i_1, l^i_2}$ correspond to the ground-state wave functions of the particle in $i^{th}$ $\R^4$ carrying angular momentum
\begin{equation}
Q_{0,i}=n_i=l_1^i+l_2^i+m^i,
\end{equation}
where $m^i$ is the intrinsic momentum with respect to the $SU(2)_R$ rotations about the origin of the $i^{th}$ $\R^4$. Thus the full field in $(\R^4)^{\otimes n}$
is 
\begin{eqnarray}
\Phi&=&\phi_1(z_{1,1},z_{1,2})\cdot\cdots\cdot\phi_N(z_{n,1}, z_{n,2})\\
&=& \sum_{\substack{l_1^1, \cdots,l^N_1 \\
                  l_2^1,\cdots,l^N_2}}\alpha_{l_1^1,l_2^1}\cdots\alpha_{l_1^N,l_2^N}(z_{1,1})^{l_1^1}(z_{1,2})^{l_2^1}\cdots(z_{N,1})^{l_1^N}(z_{N,2})^{l_2^N}.
\end{eqnarray}
Hence the total angular momentum for the mode $\alpha_{l_1^1,l_2^1}\cdots\alpha_{l_1^N,l_2^N}$ is
\begin{equation}
Q_0=\sum_{i=1}^NQ_{0,i}=\sum_{i=1}^Nn_i=\sum_{i=1}^Nl_1^i+l_2^i+m^i.
\end{equation}
The full Fock space is 
\begin{equation}
\text{Fock}|_{\text{Total}}=\underbrace{\text{Fock}|_{\R^4}\otimes\cdots\otimes\text{Fock}|_{\R^4}}_N, 
\end{equation}
where $\text{Fock}|_{\R^4}$ is the Fock space corresponding to a single $\R^4$.

Denote $\tilde{q}=e^{-g_s+iN\pi}$, where $g_s$ is the string coupling. Then the partition function is 
\begin{eqnarray}
\Z_{Fock}&:=&\text{Tr}_{\text{Total Fock}}\tilde{q}^{Q_0}Q^{Q_2} \\
&=& \text{Tr}_{\underbrace{\text{Fock}|_{\R^4}\otimes\cdots\otimes\text{Fock}|_{\R^4}}_N}\tilde{q}^{Q_0}Q^{Q_2}\\
&=&\prod_{i=1}^N\text{Tr}_{\text{Fock}|_{\R^4}}\tilde{q}^{Q_{0,i}}Q^{Q_{2,i}}\label{partition function}
\end{eqnarray}
where $Q_0=\sum_{i=1}^NQ_{0,i}=\sum_{i=1}^Nn_i$ is the total angular momentum,  $Q_2=\sum_{i=1}^N Q_{2,i}$ and $Q_{2,i}=\beta_i$ is the M2 brane charge associated to the $i^{th}$ $\R^4$. 

Following \cite{Aganagic2009}, we find that the partition function (\ref{partition function}) is
\begin{equation}
\Z_{Fock}=\prod_{\beta, m}\prod_{n=1}^\infty(1-\tilde{q}^{n+m}Q^\beta)^{nNN_\beta^m}.
\end{equation}
Here we need to note that in our case this partition function is restricted to the region $0<R \ll\infty$.
Since 
\begin{equation}
\prod_{\beta, m}\prod_{n=1}^\infty(1-\tilde{q}^{n+m}Q^\beta)^{nN_\beta^m}=\Z_{top}(\tilde{q}, Q)\Z_{top}(\tilde{q}, Q^{-1}),
\end{equation}
we have 
\begin{equation}
\Z_{Fock}=\left(\Z_{top}(\tilde{q}, Q)\Z_{top}(\tilde{q}, Q^{-1})\right)^N.
\end{equation}
Due to the wall-crossing phenomenon, we need to define the BPS partition function as a restriction of $\Z_{Fock}$ to a chamber. In our case, the total central charge is 
\begin{equation}
Z(Q_2, Q_0)=\sum_{i=1}^NZ_i(\beta_i, n_i)=\sum_{i=1}^N\frac{1}{R}(n_i+B_i\cdot\beta_i),
\end{equation}
where $B_i$ is the B field in the $i^{th}$ Calabi-Yau threefold $X$. In fact, since we place $N$ D6-branes at $N$ different centers, we have $N$ copies of the Calabi-Yau threefold $X$ placed at the $N$ centers respectively. Thus for each single D6 brane placed at the center $\vec{x}_i$ we can calculate the central charge $Z_i(\beta_i, n_i)$ of the BPS M2 brane bounded to this single D6 brane. Thus we propose that the chamber corresponding to the Coulomb phase of higher rank DT invariants is 
\begin{equation}
B_i\rightarrow \infty, \text{ for all } i=1,\cdots, N \text{ and }   0<R\ll\infty. \label{chamber}
\end{equation}
As in \cite{Aganagic2009}, in the above chamber
\begin{equation}
Z_i(\beta_i, n_i):=\frac{1}{R}(n_i+B_i\cdot\beta_i)>0, \ \ i=1,\cdots, N
\end{equation}
implies that
\begin{equation}
\beta_i>0, \ \ \text{for all }i=1, \cdots, N.
\end{equation}
So we get 
\begin{eqnarray}
&&\Z_{BPS}(B_i\rightarrow \infty, \  \  \  0<R\ll\infty)\nonumber\\
=&&\Z_{DT}^{U(1)^N}=\left(\Z_{top}(\tilde{q}, Q)\Z_{top}(\tilde{q}, Q^{-1})\right)^N|_{(\beta_i, n_i): Z_i(\beta_i, n_i|B_i\rightarrow\infty, 0<R\ll\infty )>0} \label{DT}
\end{eqnarray}

\section{Examples}
In this section we will give some examples which satisfy the proposed formula (\ref{DT}). 
\subsection{$\C^3$}
The topological string partition function for $\C^3$ is 
\begin{equation}
\Z_{top}(\tilde{q}, Q)=M(\tilde{q})^{1/2},
\end{equation}
where $M(\tilde{q})=\prod_n(1-\tilde{q}^n)^{-n}$ is the MacMahon function. So according to the equation (\ref{DT}) we have
\begin{equation}
\Z_{DT}^{U(1)^N}=(\Z_{top}(\tilde{q}, Q)\Z_{top}(\tilde{q}, Q^{-1}))^N=(M(\tilde{q})^{1/2}M(\tilde{q})^{1/2})^N=M(\tilde{q})^N,
\end{equation} 
which is same as the results in \cite{Iqbal:2003ds, Cirafici:2008sn}.

\subsection{Resolved conifold}
The topological string partition function for the resolved conifold $\calO(-1)\oplus\calO(-1)\rightarrow \bP^1$ is 
\begin{equation}
\Z_{top}(\tilde{q}, Q)=M(q)\prod_{n=1}^\infty(1-\tilde{q}^nQ)^n,
\end{equation}
which implies that the non-vanishing Gopakumar-Vafa(GV) invariants are 
\begin{equation}
N_{\beta=\pm 1}^0=1, \ \ \ N_{\beta=0}^0=-2, 
\end{equation}
and that all BPS states in five dimensions have no intrinsic spin \cite{Gopakumar1998b, Gopakumar1998c, Aganagic2009}. Thus our formula (\ref{DT}) becomes 
\begin{equation}
\Z_{DT}^{U(1)^N}=\left(\Z_{top}(\tilde{q}, Q)\Z_{top}(\tilde{q}, Q^{-1})\right)^N|_{(\beta_i, n_i): Z_i(\beta_i, n_i|B_i\rightarrow\infty, 0<R\ll\infty )>0}.
\end{equation}
According to \cite{Aganagic2009}, When $0<R\ll\infty , B_i\rightarrow\infty, i=1, \cdots, N$ we have
\begin{equation}
\Z_{top}(\tilde{q}, Q)\Z_{top}(\tilde{q}, Q^{-1})=M(\tilde{q})^2\prod_{n=1}^\infty(1-\tilde{q}^nQ)^n.
\end{equation}
So 
\begin{eqnarray}
\Z_{DT}^{U(1)^N}&=&\left(\Z_{top}(\tilde{q}, Q)\Z_{top}(\tilde{q}, Q^{-1})\right)^N|_{(\beta_i, n_i): Z_i(\beta_i, n_i|B_i\rightarrow\infty, 0<R\ll\infty )>0}\\
&=&M(\tilde{q})^{2N}\prod_{n=1}^\infty(1-\tilde{q}^nQ)^{nN},
\end{eqnarray}
which is compatible  with \cite{Cirafici:2008sn}.

\section{Conclusion and future works}
In this short paper we have investigated the relationship between M theory and the Coulomb phase of higher DT invariants and proposed a formula analogous to that of the OSV conjecture. It would be very important to check whether our proposed formula (\ref{DT}) holds in the non-toric cases. Further, it would be interesting to give the M-theory derivation of the bound states on the bound D6 branes and connect it with \cite{Nagao2010, Chuang:2010wx, Stoppa2009, Toda2009, Cirafici:2008sn} to see whether we can get Kontsevich and Soilberman's wall-crossing formula \cite{Kontsevich2008}.  Work in this direction is in progress.

\acknowledgments
The author would like to thank J. Gegenberg for valuable discussion on this project.

\bibliographystyle{JHEP}
\bibliography{MDT}

\providecommand{\href}[2]{#2}\begingroup\raggedright\begin{thebibliography}{10}

\bibitem{Strominger:1996sh}
A.~Strominger and C.~Vafa, {\it Microscopic origin of the bekenstein-hawking
  entropy},  {\em Phys. Lett.} {\bf B379} (1996) 99--104,
  [\href{http://arxiv.org/abs/hep-th/9601029}{{\tt hep-th/9601029}}].

\bibitem{Sen:2007qy}
A.~Sen, {\it Black hole entropy function, attractors and precision counting of
  microstates},  {\em Gen. Rel. Grav.} {\bf 40} (2008) 2249--2431,
  [\href{http://arxiv.org/abs/0708.1270}{{\tt arXiv:0708.1270}}].

\bibitem{Ooguri:2004zv}
H.~Ooguri, A.~Strominger, and C.~Vafa, {\it Black hole attractors and the
  topological string},  {\em Phys. Rev.} {\bf D70} (2004) 106007,
  [\href{http://arxiv.org/abs/hep-th/0405146}{{\tt hep-th/0405146}}].

\bibitem{Okounkov2007}
A.~Okounkov, N.~Reshetikhin, and C.~Vafa, {\it Quantum calabi-yau and classical
  crystals},  \href{http://arxiv.org/abs/hep-th/0309208}{{\tt hep-th/0309208}}.

\bibitem{Iqbal:2003ds}
A.~Iqbal, N.~Nekrasov, A.~Okounkov, and C.~Vafa, {\it Quantum foam and
  topological strings},  {\em JHEP} {\bf 04} (2008) 011,
  [\href{http://arxiv.org/abs/hep-th/0312022}{{\tt hep-th/0312022}}].

\bibitem{Cirafici:2008sn}
M.~Cirafici, A.~Sinkovics, and R.~J. Szabo, {\it Cohomological gauge theory,
  quiver matrix models and donaldson-thomas theory},  {\em Nucl. Phys.} {\bf
  B809} (2009) 452--518, [\href{http://arxiv.org/abs/0803.4188}{{\tt
  arXiv:0803.4188}}].

\bibitem{Sulkowski:2009rw}
P.~Sulkowski, {\it Wall-crossing, free fermions and crystal melting},
  \href{http://arxiv.org/abs/0910.5485}{{\tt arXiv:0910.5485}}.

\bibitem{Szabo:2009vw}
R.~J. Szabo, {\it Instantons, topological strings and enumerative geometry},
  \href{http://arxiv.org/abs/0912.1509}{{\tt arXiv:0912.1509}}.

\bibitem{Maulik2007}
D.~Maulik, N.~Nekrasov, A.~Okounkov, and R.~Pandharipande, {\it Gromov-witten
  theory and donaldson-thomas theory, i},
  \href{http://arxiv.org/abs/math/0312059}{{\tt math/0312059}}.

\bibitem{Maulik2007a}
D.~Maulik, N.~Nekrasov, A.~Okounkov, and R.~Pandharipande, {\it Gromov-witten
  theory and donaldson-thomas theory, ii},
  \href{http://arxiv.org/abs/math/0406092}{{\tt math/0406092}}.

\bibitem{Aganagic2009}
M.~Aganagic, H.~Ooguri, C.~Vafa, and M.~Yamazaki, {\it Wall crossing and
  m-theory},  \href{http://arxiv.org/abs/0908.1194}{{\tt arXiv:0908.1194}}.

\bibitem{Nagao2010}
K.~Nagao, {\it On higher rank donaldson-thomas invariants},
  \href{http://arxiv.org/abs/1002.3608}{{\tt arXiv:1002.3608}}.

\bibitem{Chuang:2010wx}
W.~yen Chuang, D.-E. Diaconescu, and G.~Pan, {\it Rank two adhm invariants and
  wallcrossing},  \href{http://arxiv.org/abs/1002.0579}{{\tt arXiv:1002.0579}}.

\bibitem{Stoppa2009}
J.~Stoppa, {\it D0-d6 states counting and gw invariants},
  \href{http://arxiv.org/abs/0912.2923}{{\tt arXiv:0912.2923}}.

\bibitem{Toda2009}
Y.~Toda, {\it On a computation of rank two donaldson-thomas invariants},
  \href{http://arxiv.org/abs/0912.2507}{{\tt arXiv:0912.2507}}.

\bibitem{Gaiotto:2005gf}
D.~Gaiotto, A.~Strominger, and X.~Yin, {\it New connections between 4d and 5d
  black holes},  {\em JHEP} {\bf 02} (2006) 024,
  [\href{http://arxiv.org/abs/hep-th/0503217}{{\tt hep-th/0503217}}].

\bibitem{Moore:1998et}
G.~W. Moore, N.~Nekrasov, and S.~Shatashvili, {\it D-particle bound states and
  generalized instantons},  {\em Commun. Math. Phys.} {\bf 209} (2000) 77--95,
  [\href{http://arxiv.org/abs/hep-th/9803265}{{\tt hep-th/9803265}}].

\bibitem{Jafferis:2007sg}
D.~L. Jafferis, {\it Topological quiver matrix models and quantum foam},
  \href{http://arxiv.org/abs/0705.2250}{{\tt arXiv:0705.2250}}.

\bibitem{Awata:2009dd}
H.~Awata and H.~Kanno, {\it Quiver matrix model and topological partition
  function in six dimensions},  {\em JHEP} {\bf 07} (2009) 076,
  [\href{http://arxiv.org/abs/0905.0184}{{\tt arXiv:0905.0184}}].

\bibitem{Dabholkar2008}
A.~Dabholkar, J.~Gomes, and S.~Murthy, {\it Counting all dyons in n =4 string
  theory},  \href{http://arxiv.org/abs/0803.2692}{{\tt arXiv:0803.2692}}.

\bibitem{Ida:2007vi}
D.~Ida {\em et~al.}, {\it Cosmological black holes on taub-nut space in five-
  dimensional einstein-maxwell theory},  {\em Class. Quant. Grav.} {\bf 24}
  (2007) 3141--3150, [\href{http://arxiv.org/abs/hep-th/0702148}{{\tt
  hep-th/0702148}}].

\bibitem{Witten1997b}
E.~Witten, {\it Solutions of four-dimensional field theories via m- theory},
  {\em Nucl. Phys.} {\bf B500} (1997) 3--42,
  [\href{http://arxiv.org/abs/hep-th/9703166}{{\tt hep-th/9703166}}].

\bibitem{Gopakumar1998b}
R.~Gopakumar and C.~Vafa, {\it M-theory and topological strings. i},
  \href{http://arxiv.org/abs/hep-th/9809187}{{\tt hep-th/9809187}}.

\bibitem{Gopakumar1998c}
R.~Gopakumar and C.~Vafa, {\it M-theory and topological strings. ii},
  \href{http://arxiv.org/abs/hep-th/9812127}{{\tt hep-th/9812127}}.

\bibitem{Kontsevich2008}
M.~Kontsevich and Y.~Soibelman, {\it Stability structures, motivic
  donaldson-thomas invariants and cluster transformations},
  \href{http://arxiv.org/abs/0811.2435}{{\tt arXiv:0811.2435}}.

\end{thebibliography}\endgroup
\end{document}